\documentclass{article}
\UseRawInputEncoding
\usepackage[english]{babel}
\usepackage{cite}
\usepackage{amsmath,amssymb,amsfonts}
\usepackage{algorithm, algorithmic}
\usepackage{graphicx}
\usepackage{textcomp}
\usepackage{subcaption}
\usepackage{color}
\usepackage[letterpaper,top=2cm,bottom=2cm,left=3cm,right=3cm,marginparwidth=1.75cm]{geometry}

\usepackage{amsmath}
\usepackage{graphicx}
\usepackage[colorlinks=true, allcolors=blue]{hyperref}

\title{Predicting Material Properties Using a
3D Graph Neural Network with Invariant
Representation}
\author{Boyu Zhang, Musen Zhou, Jianzhong Wu, Fuchang Gao}

\begin{document}
\maketitle

\begin{abstract}
Accurate prediction of physical properties is critical for discovering and designing novel materials.
Machine learning technologies have attracted significant attention in the materials science community for their potential for large-scale screening.
Graph Convolution Neural Network (GCNN) is one of the most successful machine learning methods because of its flexibility and effectiveness in describing 3D structural data.
Most existing GCNN models focus on the topological structure but overly simplify the three-dimensional geometric structure.
However, in materials science, the 3D-spatial distribution of atoms is crucial for determining the atomic states and interatomic forces.
This paper proposes an adaptive GCNN with a novel convolution mechanism that simultaneously models atomic interactions among all neighbor atoms in three-dimensional space.
We apply the proposed model to two distinctly challenging problems on predicting material properties.
The first is Henry's constant for gas adsorption in Metal-Organic Frameworks (MOFs), which is notoriously difficult because of its high sensitivity to atomic configurations.
The second is the ion conductivity in solid-state crystal materials, which is difficult because of few labeled data available for training.
The new model outperforms existing graph-based models on both data sets, suggesting that the critical three-dimensional geometric information is indeed captured.
\end{abstract}

\section{Introduction}
\label{sec:introduction}
Accurate and rapid prediction of physical properties plays an essential role in discovering and designing novel materials.
\textit{Ab initio} methods \cite{simunek2006hardness} \cite{soler2002siesta} are commonly used to predict the properties of a material.
However, \textit{ab initio} methods, such as Density-Functional Theory (DFT) \cite{obot2015density} and wave function theories \cite{booth2013towards}, are time-consuming.
An alternative method is to predict these properties using machine learning (ML) technologies. 
By modeling the material structure and atomic interactions, ML algorithms are much faster than DFT calculations  and yet provide similar accuracy \cite{RN34}.

Although ML techniques have succeeded in many fields, such as computer vision and natural language processing, materials science is still significantly challenging \cite{wei2019machine}.
The atomic structures of the materials are complex three-dimensional spatial structures with an unfixed number of atoms.
The classic machine learning models, such as Convolutional Neural Network (CNN), are better at processing fixed-size data.
In contrast, graph-based models, such as Message Passing Neural Network (MPNN) \cite{gilmer2017neural} and Graph Convolutional Neural Networks (GCNN), have the advantage of flexibility in describing variable data.
Moreover, the message passing (for MPNN) and convolution (for GCNN) could effectively stimulate the atomic interaction in the materials. 
Recently, graph-based models have attracted significant attention because of the flexibility and effectiveness in describing 3D structures formed by discrete particles \cite{allam2018application, RN29, Schutt2018schnet, xie2018crystal}.

Although the effectiveness of the model has been well demonstrated, the current GCNN still has some shortcomings.
GCNN describes the material structures with undirected graphs, where nodes represent the atoms, while edges express the relationship between bonded atoms. 
The graph-based description is concise and efficient but ignores the material's three-dimensional topology, which critically impacts material properties.
In addition, the GCNN updates the state of each atom by accumulating the impact of the surrounding bonded atoms. This design is based on the assumption that the surrounding atoms independently influence the current atom. 
However, the surrounding atoms exert their influence on the current atom simultaneously.
The model would be more effective if it were true to reality \cite{greydanus2019hamiltonian}.

This paper presents k-Nearest Atoms Graph Convolution Network (k-NAGCN) for predicting the physical properties of materials.
The proposed model improves the accuracy of attribute prediction by introducing three-dimensional topological structure information. 
We also designed an innovative convolutional structure to simultaneously calculate the influence of surrounding atoms on the current atom.
Besides, we developed an algorithm to calculate translate and rotation invariant descriptors for the local atomic distribution.
We tested the new model with two distinctly challenging problems on predicting material properties.
The first is Henry's constant for gas adsorption in Metal-Organic Frameworks (MOFs), which is notoriously difficult because the results are highly sensitive to inter-atomic separations.
The second example is the prediction of ion conductivity in solid-state crystal materials. The task is also difficult due to the lack of labeled data available for training.
The k-NAGCN model presents superior performance on both datasets, suggesting that it indeed captures important three-dimensional geometric information.

The contributions of this paper are as follows.
\begin{itemize}
  \item Proposed an end-to-end system for predicting the physical properties of materials.
  \item Modeled the three-dimensional material structures with a graph-based model and proposed a novel convolution mechanism that better simulates physical laws.
  \item Developed a rotation- and translation-invariant descriptor to model the local topology of atoms.
\end{itemize}

\section{Related Work}
{\color{black}
The emergence of large-scale datasets of materials\cite{hellenbrandt2004inorganic, jain2013commentary, kirklin2015open} greatly empowers the applications of novel machine learning technologies in the field of materials science \cite{RN6}.}
The representation of materials plays a critical role in building an effective machine learning system, and it is challenging due to the complexity of the chemical structures.
{\color{black}
One strategy is to describe the materials using a group of physical properties.
Cubuk et al.\cite{cubuk2019screening} trained a support vector machine (SVM) based on thirty features and then used transfer learning to get a formula-based predictor. The formula-based predictor offered significant advantages in efficiency, being able to screen billions of materials in a short period.
Allam et al. \cite{allam2018application} employed DFT modeling approach to predict basic quantum mechanical quantities. The predicted features were combined with the structure feature to obtain a neural network to predict molecular electrode materials.}
{\color{black}
Feature-based methods rely on expertise in materials science, and their performance also depends on the effectiveness of the features used.
}

{\color{black}
With large-scale data available, researchers tend to use data-driven approaches that make greater use of low-level structural information.}
{\color{black} Hoffmann et al.} \cite{hoffmann2019data}, for instance, constructed a smooth and continuous three-dimensional density representation of each crystal based on different atomic positions and designed an {\color{black}autoencoder/decoder network to predict innovative materials.}
{\color{black} The method benefits from the detailed structural information but requires high computational power.}
{\color{black} Zheng et al.}\cite{zheng2018machine} presented the atomic configuration as a two-dimensional periodic table and normalized the table to mimic a digital image.
They trained a multitask Convolutional Neural Networks (CNN) model that outputs the lattice parameters and enthalpy of formation simultaneously.
The experimental results supported that their method achieved competitive performance on ICSD \cite{hellenbrandt2004inorganic}  and OQMD \cite{kirklin2015open}.
{\color{black}Schütt et al. \cite{Schutt2018schnet} also described materials using periodic tables. The proposed architecture, Schnet, used a continuous filter convolutional layers to model local correlations and yielded excellent performance for predicting a group of molecular properties.}

Graph-based models could handle variant inputs and thus have attracted much attention in chemistry.
{\color{black}Xie et al. \cite{xie2018crystal} proposed Crystal Graph Convolutional Neural Network (CGCNN) that described crystals using undirected graphs, where the vertices and edges corresponded to the atoms and bonds, respectively.
CGCNN captured the atomic interactions using the graph convolution layers, and a pooling layer embedded the graph into the feature space.}
The CGCNN model showed excellent performance on various properties, including formation energy, bandgap, Fermi energy, and elastic properties.
{\color{black} Sanyal et al.} \cite{RN16} adapted the original CGCNN using multitask learning \cite{RN17}, and their experiments showed that sharing parameters in the lower level improved the prediction performance.

{\color{black}Domain knowledge can be integrated into data-driven models to improve performance.}
\cite{RN11} developed a universal model to predict the physical properties for both molecules and crystals by incorporating state variables, such as temperature, pressure, and entropy.
\cite{RN19} added a relative position matrix into the model and provided a mechanism to handle the feature update process.
The proposed model, 3DGCN, showed good performance for predicting molecular properties and biochemical activities.

\section{Method}
{\color{black}This section explains the details of the proposed k-NAGCN. 
The model accepts the structure of the material as input. 
After modeling the structure, convolution operations are performed on the graph, and finally, the features of each node are aggregated, and the target attributes are estimated.
The following three subsections discuss graph-based structure representation, convolution operations, and the algorithm that produces translation- and rotation-invariant structural descriptors in three-dimensional space, respectively.
}

\begin{figure*}[!h]
    \centering
    \begin{subfigure}[b]{0.75\textwidth}
    \includegraphics[width=1\linewidth]{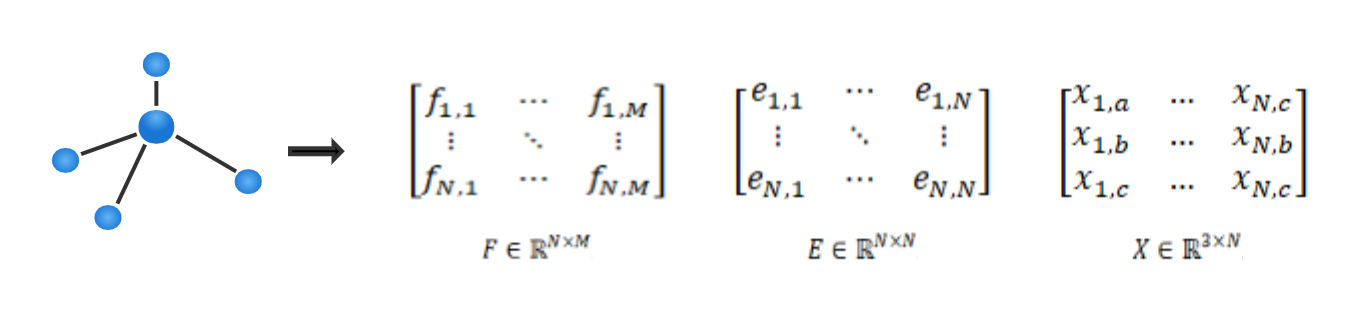}
    \caption{Graph-based representation.}\label{method_representation}
    \end{subfigure}
	
	\begin{subfigure}[b]{0.55\textwidth}
    \includegraphics[width=1\linewidth]{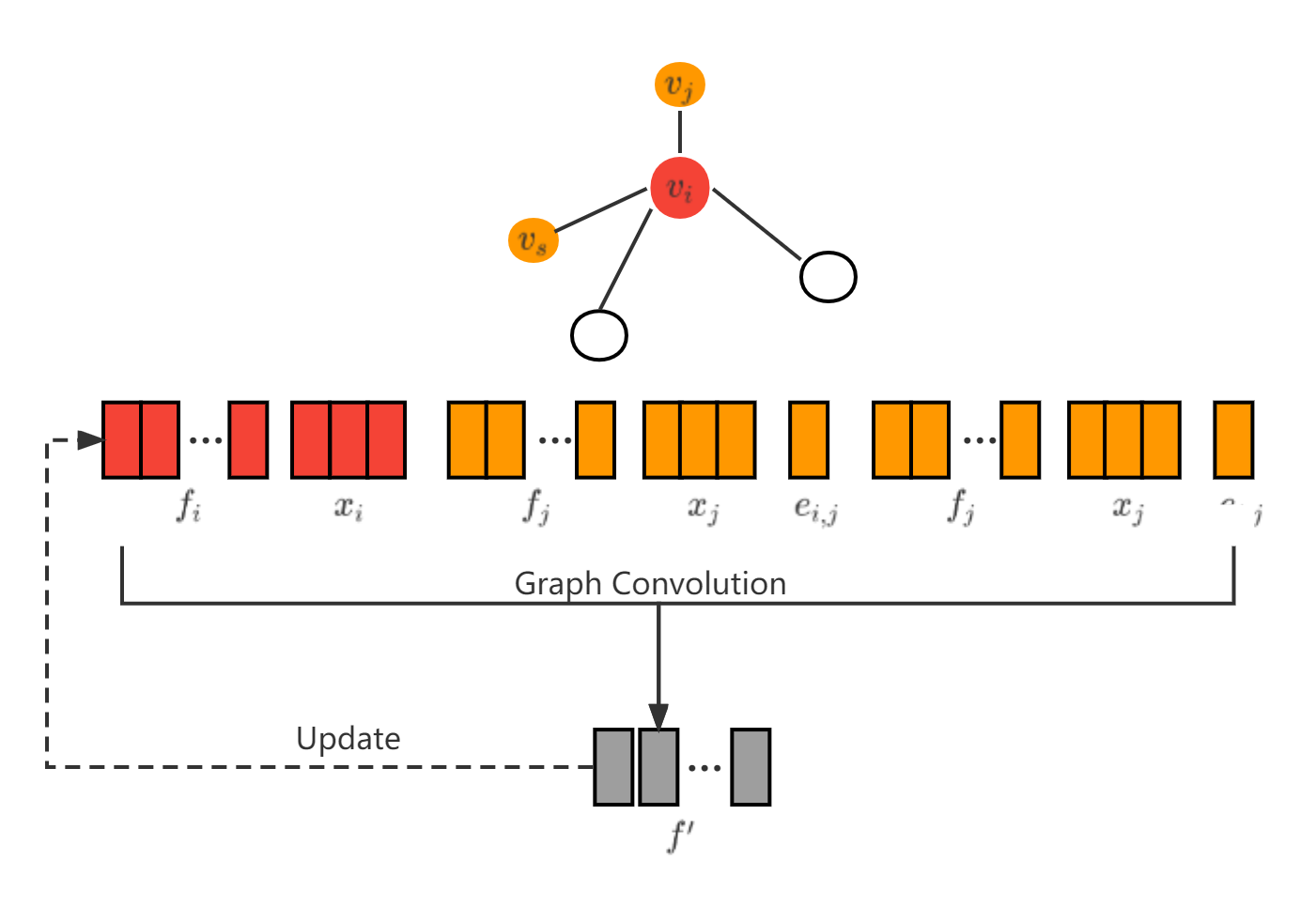}
    \caption{The process of convolution in k-NAGCN}\label{method_convolution}
    \end{subfigure}
	
    \caption[]{(a) The input material (left) is encoded into three pieces of information, the atomic features $F$, the edge features $E$, and the spatial locations $X$. (b) Based on the representation in (a), the convolution is calculated as follows ($k=2$). For vertex $v_i$, the two closest neighbor atoms are $v_j$ and $v_s$. The atomic features ($f_i, f_j, f_s$), the edge feature ($e_{i,j}, e_{i,j}$, and the spatial locations are concatenated and used as the input of the convolution and the output, $f'$ is accumulated with the original feature $f_i$.}
\end{figure*}

\subsection{Graph-based Representation}\label{description}

% Effective and concise representation of atomic structures of materials is a challenging topic in the field, and a variety of methods have been proposed \cite{RN30}.
% Among these methods, graph-based methods present excellent computational efficiency and have superiority in modeling the inner complexity of the structures.
% Models based on GCN have been successfully applied to property prediction for crystals, and molecules \cite{xie2018crystal, RN16} , and the results demonstrated that these models are effective in general.

{\color{black}This subsection presents the graph-based material model that considers the three-dimensional topology.
A material contains $N$ atoms is represented by an undirected graph $G=(V,E)$, where $V = \{ {v_1},...{v_N}\} $ is the set of vertices and $E = \{ {e_{i,j}}{\rm{|}}i,j \in {\rm{\{ }}1,...N{\rm{\} }},i \ne j\} $ is the set of edges.
The information associated with the graph includes three pieces, the atomic feature $F$, the edge feature $E$, and the coordination $X$ (see Figure \ref{method_representation}).}

{\color{black}Each node in the graph $G$ is associated with a group of features that reflect the prior chemistry knowledge about the elements.}
In this article, an $M \times L$ matrix  ${M_f}$ was calculated using the adaptive `atom2vector' method \cite{RN31}, where $M$ is a hyper-parameter that reflects the dimension of the atom feature, and $L$ is the total number of different elements in the dataset.
Each column of ${M_f}$ is an $M$-dimensional feature vector of the corresponding element.
${M_f}$ summarizes the domain knowledge about the elements in the dataset and is independent of the material structures.
{\color{black}One-hot encoding \cite{RN35} is used to represent the element type. 
$A$ is an $N$ by $L$ matrix, where $a_{i,j}$ equals to $1$ if the $i$-th atom in the input structure has the element type $j$, and $0$ otherwise.
For an input material, the feature matrix $F$ is calculated using $A \times M_f$.
The $i$th row of $F$, $f_i$, is the atomic feature associated with the $i$-th vertex.}
{\color{black}The spatial distribution of the atoms is represented by the coordinate matrix $X$, which is  $3\times N$, and the $i$-th column is the atomic positions corresponding to the unit cell dimensions.

In most graph-based representations, the adjacency matrix is used to describe the chemical bonds. 
This paper replaces the adjacency matrix by the distance matrix because k-NAGCN calculates the graph convolution over each atom and its $k$ nearest neighbors, bonded or not bonded. 
Element $e_{i,j}$ in matrix $E$ represents the distance between atom $i$ and $j$.}

% aim to find an optimized function F that minimizes the distance between ground truth $p$ and the prediction $\hat p = F({M_f},A,X)$, where $\hat p$ is the predicted value of the property of interest.

\subsection{k-Nearest Neighbor Convolution} \label{convolution}
{\color{black}The graph convolution has the most significant impact on the performance of graph convolution neural networks.
This subsection explains the proposed convolution mechanism that involves the nearest neighbor atoms simultaneously.}

{\color{black}With the material represented by graph $G=\{V,E\}$, each vertex $v_i$ is associated with the atomic feature $f_i$ and coordination $x_i$ and each edge is associated with the edge feature $e_{i,j}$.
The distance $e_{i,j}$ is expand using a series of pre-chosen numbers,  ${\alpha_1}, {\alpha_2},,…, {\alpha_{32}}$. 
Thus each edge is associated with a edge feature vector $({e^{ - {\alpha _1}\|{e_{i,j}}\|}}, {e^{ - {\alpha _2}\|{e_{i,j}}\|}},…, {e^{ - {\alpha _32}\|{e_{i,j}}\|}})$.
or simplicity, the next part of this paper does not distinguish the scalar $e_{i,j}$ and vector $e_{i,j}$.}

{\color{black}The atomic feature $f_i$ associated with the $i$-th atom is updated according to equation (\ref{newconv}).
\begin{equation}\label{newconv}
\hat {f_i} = f_i + Conv(f_i,\cup_{j \in K_i}(e_{i,j},f_j,{x_j^{out}}))
\end{equation}
where $Conv(\cdot)$ is a combination of the linear transformation and nonlinear activation function, and $K_i$ is the set $k$ indices of the $k$ nearest vertices that are closest to the $i$-th vertex.
$x^{out}$ is the transformed coordination calculated using Algorithm \ref{atomsift}.}
$k$ is a hyper-parameter that determines how many neighbor atoms should be involved in the convolution based on the distance.
Figure \ref{method_convolution} presents the calculation of k-NAGCN convolution used by k-NAGCN when $k=2$. 
The proposed convolution considers the impacts of $k$ nearest neighbor atoms simultaneously, and thus, is a better simulation of real-world atomic interactions.

\subsection{Rotational Invariant Local Descriptor}
{\color{black}The convolution proposed in subsection \ref{convolution} is sensitive to the order of atoms and translation.
This subsection proposed translation and rotation invariant descriptors for the local atomic distribution.}
% There is one significant difference between the convolution mechanism described in equation (\ref{oldconv}) and (\ref{newconv}).
% The former is independent of the order of the vertices because it sums up all the pairwise interactions between atoms.
% By contrast, the latter mechanism is order-sensitive because of the spatial information.

{\color{black}It is vital to keep the atom sequence identical when the local structure rotates or translates, which are very common in materials.}
Assuming $k=2$ and the two neighbor vertices of vertex $v_i$ is noted as $v_j$ and $v_k$.
Then the input of $Conv(\cdot)$ is $(f_i,X_i,e_{i,j},f_j,X_j,e_{i,k},f_k,X_k)$.
Use $v_i$ as the center, we could rotate the geometric composed of $v_i$, $v_j$, and $v_k$ until $v_j$ and $v_k$ switch the place with each other.
Then the input of $Conv(\cdot)$ becomes $(f_i,X_i,f_{i,k}, f_k,X_k,f_{i,j},f_j,X_j)$ and the output is expected to change accordingly.
However, the chemical feature of atom $v_i$ does not change when rotated.

% Theoretically, the model can learn the rotated coordinates' equivalency if a large group of training samples is provided.
% Unfortunately, we do not have enough training samples under most circumstances, and it is computationally expensive.
% This paper developed a simple and effective method to achieve rotational invariant for the atom sequences.

% There are existing researches devoted to invariant descriptions.
% Most of these descriptions have a high computational cost and could not fit deep neural network structures.
% The scale-invariant feature transform (SIFT) \cite{RN21} is the most widely used algorithm that generates rotation invariant descriptors for local patches in images.
% For each located keypoint, the SIFT descriptor calculates the gradient distribution and decides the principal direction.
% Then the algorithm rotates the principal direction to the x-axis, and the description vector was calculated.
% The SIFT algorithm has been extensively investigated and applied widely.
% There are a few adaptions of the original SIFT algorithm \cite{RN22, RN23, RN24} that extended the algorithm to 3D space.
% Comparing to images, the calculation in 3D space is more complex due to the additional degree of freedom.
% However, the chemical structure data under our consideration are discrete and sparse, and thus, the calculation could be significantly simplified.
Inspired by the Scale-Invariant Feature Transform (SIFT) \cite{RN21}, which is widely used in the field of computer vision, We proposed an algorithm that output identical atom sequence and coordinates that are invariant to translation and rotation (Algorithm \ref{atomsift}).

\begin{algorithm}
 \caption{The rotational invariant description for local structure} \label{atomsift}
 \begin{algorithmic}[1]
 \renewcommand{\algorithmicrequire}{\textbf{Input:}}
 \renewcommand{\algorithmicensure}{\textbf{Output:}}
 \REQUIRE Graph $G=\{V, E\}$ ,  coordinate matrix $X$, number of neighbor atoms $k$  
 \ENSURE  Transformed coordinate matrix $X_{out}$
 \\ \textit{LOOP Process}
  \FOR {$i\leftarrow 1$ to $N$}
  \STATE Calculate distance between vertex $v_i$ and $v_j$, where $v_j \in V, \neq j$\;
  \STATE Sort $v_j$ based on distance and get $k$ nearest neighbors \;
  \STATE Transform the origin of coordinate system to $x_i$\;
  \STATE Calculate transform matrix $R_1$ that rotates $x_1^i$ to $[0,0,|x_1^i |]^T$ using Rodrigues' rotation formula\;
  \STATE Update the coordinate by $X_{R_1}^i=R_1 \times X^i$\;
  \STATE Calculate transform matrix $R_2$ that rotate $x_2^i$ to $yz$-plane\;
  \STATE Update the coordinate by $X_{out}^i=R_2 \times X_{R_1}^i$
  \ENDFOR
 \RETURN $X_{out}$ 
 \end{algorithmic} 
 \end{algorithm}

For each atom $v_i$ in the structure, we calculated the distance between $v_i$ and the rest atoms $\{v_j\}$, where $ i \neq j$.
Then we sorted all $v_j$ according to the distance and found the $k$ nearest atoms based on the hyper-parameter $k$.
Denote by $x_i$ the position of atom $v_i$, and by $x_i^1$ and $x_i^2$ the positions of the closest and the second closest atoms to $x_i$.
We first find transformation $T$ that translates $x_i$ to the origin.
Next, we calculate a matrix $R_1$ that rotates $x_i^1$ to positive $z$-axis according to Rodrigues' rotation formula \cite{RN33}.
Then, we calculated the matrix $R_2$ that rotate $x_2^i$ to the $yz$-plane, keeping $x_i^1$ fixed.
Finally, we calculated the new coordinate $R_2 \circ R_1 \circ T(x)$ for every atom in the $k$ nearest neighborhood.
It is readily seen that this distance-based sort and rotation guarantee the rotational invariance.

\section{Experiment}

\subsection{Experimental Data}
The proposed k-NAGCN was tested on two experimental datasets.
% 10136
The first dataset, CoRE2019, contains 12763 MOFs, and the property of interest is Henry's constant \cite{RN26}.
Appendix \ref{appendix:Henry} presents the details of the calculation of Henry's constant.

The proposed algorithm was also tested with a subset of the ICSD database built for ion conductivity prediction.
The properties of interest of the ICSD data include minimum bond length, large Li site, and percolation radius.
{\color{black}In detail, there are 1758 crystals available for the bond length prediction, 2818 crystals available for percolation radius prediction, and 894 crystals available for Li site prediction.}
These properties were calculated through topological analysis and could be used as indicators for high potential conductors. 
Readers could refer to \cite{RN25} for more details of the calculation of these properties.

\subsection{Implementation details}
{\color{black}
\subsubsection{Data cleaning for CoRE2019.} The original CoRE2019 MOF database contains 12763 MOF structures, and the range of the Henry's constant covers more than 30 orders of magnitude.
We examined the structure files and found the broad range of Henry's constant was caused by unreasonable structures.
In a small portion of MOFs, atoms were unreasonably close to each other, and the overlapped atoms led to high attraction and high Henry's constant.
Meanwhile, there were another group of MOFs that had tightened confined geometry, and the thin structures resulted in extremely low Henry's constant(at the order of $10^{-10}$).

We got rid of MOFs with atomic distance smaller than $0.8$ $\mathring{A}$ to avoid overlapped atoms. 
We also filtered out the MOFs with extremely low Henry's constant by removing structures with a surface area less than 50 $\mathring{A}^2$ (meaning that there will be very little space and very unlikely for surface adsorption to happen).
The cleaned dataset contains 10136 MOFs.
The details of the above sample selection process could be found in Appendix \ref{appendix:sample}.

\subsubsection{Value Transformation of Henry's Constant} The Henry’s constants still covered several orders of magnitudes. 
During the backpropagation of the gradient, the tremendous values would cause gradient explosions and make the training infeasible.
Existing research (\cite{RN27} predicted the value $\log_{10}(H_{cc})$ instead of the widely varied original value.

In this paper, we transformed the original value to $\log_{10}(H_{cc})$ and $\log_{10}(H_{cc} +1)$.
The two terms were close with large $H_{cc}$, but the latter term put less weight on materials with low Henry’s constants.
The advantage of the latter term $\log_{10}(H_{cc} +1)$ was that materials with high Henry's constant were favored for gas adsorption, while materials with extremely low Henry's constant was not of the practical interest.

\subsubsection{Experimental setup and parameter selection} 
10-folded cross-validation was used through the experiments to ensure the fairness of evaluation.
In detail, samples were randomly divided into training set $80\%$ validation set $10\%$, and testing set $10\%$.
The network was trained using the training set, and the performance was evaluated on the testing set.
The validation set was used for the selection of hyperparameters, such as numbers of training epochs and convolution layers, during the training process.
The process was repeated ten times and the performance was evaluated using the average of the Mean Absolute Error (MAE). 

The following parameters were shared for all tasks.
The batch size is set as $32$ to avoid overfitting.
Adam is used as the optimizer to find optimized parameters. 
The initial learning rate was set as $0.0003$. The learning rate was reduced by a factor of $0.1$ for every $50$ epoch.

The experiments were performed on a Microsoft Azure virtual machine with 24 vCPU (2.5GHz), 112 G ram, and 1 V100 GPU (5120 CUDA Cores and 32Gb ram).
}

\subsection{Results}
{\color{black}We first predicted $H_{cc}$ of the CoRE2019.}
{\color{black}All 10136 MOFs were used in the experiments.}
Both $\log_{10}(H_{cc})$ and $\log_{10}(H_{cc}+1)$ were predicted.
{\color{black}The performance were compared with two graph-based models, CGCNN \cite{xie2018crystal} and 3DGCN \cite{RN19}, which were designed for material property prediction.}

{\color{black}Both CGCNN and 3DGCN were implemented according to the original literature and adapted to fit the CoRE2019 data.}
For CGCNN, the network structure includes one feature embedding layer, three graph convolution layers, and three fully-connected layers based on the performance of the validation data.
For the 3DGCN, max pooling was used because the original literature reported max pooling outperformed average pooling.
{\color{black}The k-NAGCN contained three graph convolution layers and two fully-connected layers, and average pooling is used.}
{\color{black}10-folded cross-validation was used for all three models.}
The results are summarized in Table \ref{tb_prediction}.

\begin{table}[t]
\centering
\caption{Average MAE of Henry's constant prediction for CoRE2019 dataset.}\label{tb_prediction}
\setlength{\tabcolsep}{3.5mm}{
\begin{tabular}{*8c}
\hline
\rule{0pt}{3ex} &  \multicolumn{2}{c}{$log_{10}(Hcc)$} & \multicolumn{2}{c}{$log_{10}(Hcc+1)$}\\
\cline{2-5}
\rule{0pt}{3ex} & MAE & STD & MAE & STD \\
\hline
\\[-0.5em]
CGCNN \cite{xie2018crystal}    & 0.31 & 0.07 & 0.25  & 0.05 \\
\\[-0.5em]
3DGCN \cite{RN19} & 0.22 & 0.04 & 0.16 & 0.04 \\
\\[-0.5em]
k-NAGCN & \textbf{0.15} & 0.04 & \textbf{0.10} & 0.03
\end{tabular}}
\end{table}

\begin{figure*}
        \centering
        \begin{subfigure}[b]{0.475\textwidth}
            \centering
            \includegraphics[width=\textwidth]{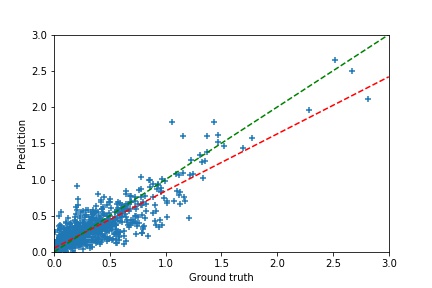}
            \caption[]%
            {{\small Henry's constant (MOFs)}}    
            \label{scatter_hcc}
        \end{subfigure}
        \hfill
        \begin{subfigure}[b]{0.475\textwidth}  
            \centering 
            \includegraphics[width=\textwidth]{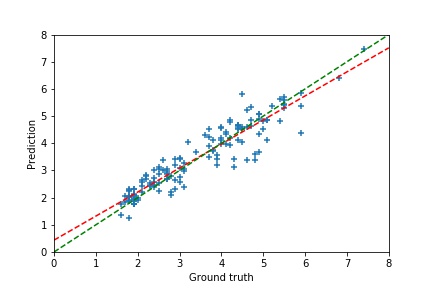}
            \caption[]%
            {{\small Minimum bond length} }    
            \label{sactter_bond}
        \end{subfigure}
        \vskip\baselineskip
        \begin{subfigure}[b]{0.475\textwidth}   
            \centering 
            \includegraphics[width=\textwidth]{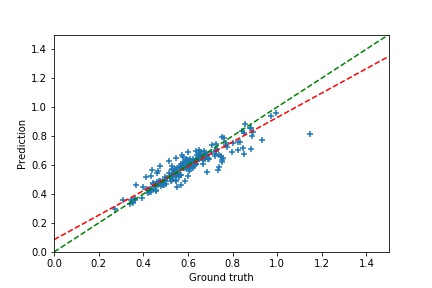}
            \caption[]%
            {{\small Percolation radius (ICSD)}}    
            \label{scatter_radius}
        \end{subfigure}
        \hfill
        \begin{subfigure}[b]{0.475\textwidth}   
            \centering 
            \includegraphics[width=\textwidth]{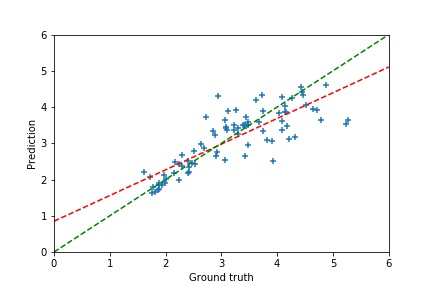}
            \caption[]%
            {{\small Li site}}    
            \label{scatter_site}
        \end{subfigure}
        \caption[ The average and standard deviation of critical parameters ]
        {\small The prediction results of material properties use k-NAGCN. The horizontal coordinate indicates the ground truth values and the vertical coordinate indicates the predicted values. The red dashed lines are the the trend lines for each predicted set, and the green dashed lines indicate the identity lines. (a) Henry's constant of MOFs. (b) Minimum bond length of ICSD. (c) Percolation radius of ICSD. (d) Large Li site of ICSD. } 
        \label{scatter}
    \end{figure*}

{\color{black}k-NAGCN outperformed CGCNN and 3DGCN on $H_{cc}$ prediction of CoRE2019.
This result proved the two hypotheses we made in the previous section. First, 3DGCN and k-NAGCN achieved better results in $H_{cc}$ prediction, proving that 3D topology had an essential role in property prediction. Second, k-NAGCN with the new convolution mechanism achieved lower MAE than 3DGCN, proving that the proposed convolution mechanism could better model realistic atomic interactions. 
Figure \ref{scatter_hcc} gives the scatterplot of the prediction results of $H_{cc}$ on Core2019 dataset.
}

{\color{black}All three approaches achieved smaller MAE on $\log_{10}(H_{cc}+1)$ than $\log_{10}(H_{cc})$.
The result was predictable because $\log_{10}(H_{cc}+1)$ had smaller variation compared with $\log_{10}(H_{cc})$.}
Considering both terms are one-to-one mapping, the $\log_{10}(H_{cc}+1)$ is a better choice for $H_{cc}$ prediction task.

\begin{figure}
	\centering
	\includegraphics[scale=0.5]{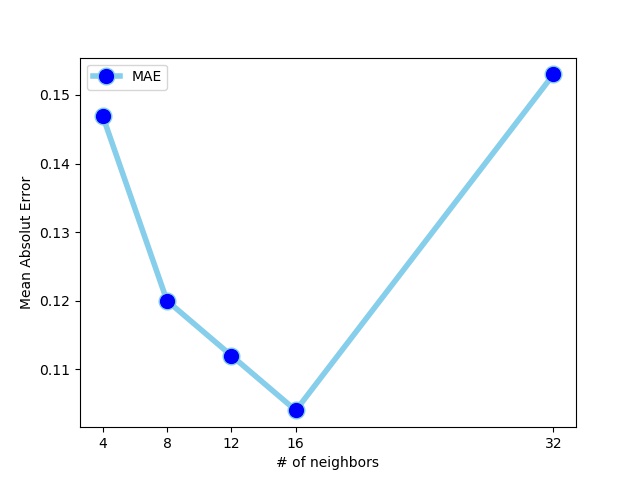}
	\caption{The effect of the number of neighbors in the convolution. Presented on the prediction of $\log_{10}(1+H_{cc})$.}
	\label{keffect}
\end{figure}

One crucial hyper-parameter was the number of neighbor atoms $k$ while calculating the convolution (see equation \ref{newconv}).
{\color{black}$k$ was important because it determined the scale of the local structure, similar to the convolutional kernel size in a CNN.}
To evaluate the impact of the $k$, we repeated the experiments using the same setup with different scale parameters, which are 4, 8, 12, 16, and 32.
The best average MAE achieved using different $k$ was summarized in Figure \ref{keffect}.

The number of neighbor atoms had significant impacts on the performance.
The MAE is significantly larger when only a few neighbor atoms, e.g., four, were involved in the convolution.
When $k$ was too small, the convolution did not consider the target atom's whole local environment, and the model underperformed.
The MAE was reduced while increasing the number and putting more atoms into the calculation.
However, when we increased the size of the neighborhood to a specific range (e.g., 12 and 16), a further increment did not increase the performance.
{\color{black}The performance decreased significantly when we increased the $k$ to 32.
There were two reasons for the deterioration. 
First, the convergence was affected by the increased parameter number. 
Secondly, when the scale was too large, the atomic features were over-smoothed.}

The 3DGCN, which also considered the 3D topology, performed slightly better than the CGCNN but lower than k-NAGCN.
The 3DGCN modeled the atomic interactions with pairwise additive interactions, but the surrounding atoms simultaneously interact with the target atom.
The experimental results prove that modeling all neighbor atoms at one time better fitted the physics laws.

We also tested the new algorithm on the ICSD crystal data.
The performance was evaluated using MAE, and the results are summarized in Table \ref{tbl2}.

\begin{table*}[!htbp]
\centering
\caption{Performance of the proposed algorithm on ICSD data.}\label{tbl2}
\setlength{\tabcolsep}{.9mm}{
\begin{tabular}{*8c}
Database & \# of training data & Unit & CGCNN \cite{xie2018crystal} & 3DGCN \cite{RN19} & k-NAGCN\\
\hline
Large Li-Site                & 863 & $\mathring{A}$ (0.1 nm) & 0.390 & 0.375 & \textbf{0.329} \\
\hline
Percolation Radius       & 2524 & $\mathring{A}$ (0.1 nm) & 0.063 & 0.060 & \textbf{0.052} \\
\hline
Minimum Bond Length  & 1760 & $\mathring{A}$ (0.1 nm) & 0.697 & 0.700 &  \textbf{0.653}\\
\end{tabular}}
\end{table*}

The results in table \ref{tbl2} present that the proposed method outperformed two graph-based methods on predicting topology features.
According to \cite{RN25}, the enlarged Li site originates from the large local space within the crystal structural framework, and the percolation radius is defined as the maximum radius of a sphere that can percolate across at least one direction of the structure.
Both definitions involve the three-dimensional topology information, and the proposed algorithm benefited from the extra information and thus outperformed the existing model.
The ICSD data set experiments validated our hypotheses that introducing three-dimensional information into the graph-based representation improves the model performance.
Figure \ref{scatter} presents the prediction results using k-NAGCN.

One concern about the proposed method was the computational cost.
There were two important parts when discussing computational efficiency, graph building, and convolution calculation.
Our algorithm added extra steps to generate rotation invariant coordination, which cost more time.
Meanwhile, the new convolution layer had more parameters than the existing methods.
Assuming the convolution involves four neighbor atoms, and the atom feature and the edge feature are 128 and 32, respectively.
The new convolution had slightly over 100 thousand parameters versus 36 thousand parameters for the convolution in CGCNN \cite{xie2018crystal}.
The average times consumed for samples used in the experiments were summarized in Table \ref{tbl1}.

\begin{table}[!htbp]
\centering
\caption{Average time consumed for processing samples (ms).}\label{tbl1}
\setlength{\tabcolsep}{2mm}{

\begin{tabular}{*8c}
\hline
\rule{0pt}{3ex} &  \multicolumn{2}{c}{Graph Building} & \multicolumn{2}{c}{Training}\\
\cline{2-5}
\rule{0pt}{3ex} & k-NAGCN & CGCNN & k-NAGCN & CGCNN \\
\hline
\\[-0.5em]
Core2019 \cite{xie2018crystal}    & 493.15 & 335.89 & 1.68  & 3.99 \\
\\[-0.5em]
ICSD \cite{RN19} & 120.7 & 113.06 & 1.31 &1.54 \\
\end{tabular}}
\end{table}

As expected, it could be found that the extra process during graph building increased the computational cost.
For MOFs data, the average processing time increased from 335.89 milliseconds to 493.15 milliseconds for the graph building and from 113.06 milliseconds to 120.7 milliseconds for the forward process.
It is noticeable that the computational cost of training (forward and backward) reduced significantly.
The reason is that the method in \cite{xie2018crystal} needed to sum up the impacts of the neighbor atoms, and multiple convolutions were performed.
Improvement could be made by introducing more computation units and computing the convolutions in parallel.
However, the possible improvement is by no means without limitations.
Compared to existing algorithms, the proposed method improved the overall computational efficiency by reducing the training time.

\section{Discussion}
Accurate and rapid prediction of the physical properties of materials is a challenging problem.
The recent major advance of deep learning and the emergence of large-scale material databases empower the development of machine learning technologies in the field.
However, the variant material structure still stands in the way, hindering the immense success of these technologies.
The graph-based representation is concise and helpful in describing variant structures and effectively handling most materials analysis tasks.
The undirected graph could hardly carry the higher-dimensional topology, which is vital to understanding the atoms' interaction.
This paper develops the k-NAGCN model that involves three-dimensional topological information.

The three-dimensional information makes the process more complex than the operation on the graph.
One of the core problems is that the operation should be rotation invariant in the three-dimensional space.
Inspired by the classic SIFT algorithm, we propose to rotate a new description that is rotation invariant.
Moreover, Both \cite{RN19} and \cite{hoffmann2019data} reported reduced calculation efficiency after involving the 3D information into consideration.
The additional computational cost of the new convolution depends on the total number of atoms in the structure and the neighborhood scales $s$, which is a constant.
Thus, the computational complexity for description calculation is $O(s \times n)$.
The above computational cost does not affect the forward/backward process and has little impact on the training process.

The concept of scale is one significant difference between the existing model and the new model.
Most existing methods loop over all the chemically bonded atom pairs and sum up the atomic interaction.
The k-NAGCN eliminates the connected matrix and considers the proper scale parameter to prospect the corresponding local structure.
One may question about the fixed scale parameter would limit the flexibility of the model. 
However, the atomic interaction can be learned from data by considering the atom features and 3D atomic configuration, and it is unnecessary to be predefined.

\section{Conclusion and Future Work}
This work investigated three-dimensional geometric information usage in the graph-based model and proposed a novel k-NAGCN model.
By comparing the experimental results, we found that 3D geometry played a vital role in modeling the effects of interatomic interactions on the physical properties of materials.  The proposed network structure better reflected real-world interaction compared to existing graph-based models that mainly ignore 3D topology.

Meanwhile, we developed an algorithm that generated rotation invariant descriptors for the distribution of local atoms.
The invariant descriptor enhances the generalization of the prediction model by eliminating the ambiguity caused by rotation.
Due to the benefits brought by our improvements, the k-NAGCN model outperformed the existing methods on evaluating Henry's constants for gas adsorption in MOFs and crystal property predictions.

Our experiments proved that modeling materials structure as a 3D point cloud is feasible.
The proposed model could be extended to general tasks for predicting the physical properties of materials and could be easily embedded in other models such as graph auto-encoders and graph generative networks.
Meanwhile, introducing the scale parameter helps discover the most significant local environment for the atoms and may lead to novel meaningful chemistry structures.

The future work involves two parts. On the one hand, the proposed model will be used and validated on a series of property prediction tasks. 
It will also be used as the discriminator in the Generative Adversarial Network (GAN) that generates new MOFs. 
On the other hand, algorithms will be developed to enhance interpretability, which prompts the discovery of physic laws.

\bibliographystyle{alpha}
\bibliography{henry}

\clearpage
\appendix

\renewcommand\thefigure{\Alph{section}\arabic{figure}}
\renewcommand\theequation{\Alph{section}\arabic{equation}}

\section{The calculation of Henry's Constant}\label{appendix:Henry}

Here, we present the Henry's constant in the dimensionless format calculated as followings \cite{RN26},
\setcounter{equation}{0}%%
\begin{equation}
{H^{*}_{cc}} = \frac{1}{{V}}\int {{e^{ - \beta {V^{ext}}({\bf{r}})}}d{\bf{r}}}
\end{equation}
where $\beta={1}/{(k_BT)}$, $R$ represents the gas constant, $k_B$ stands for the Boltzmann constant.
$T$ is the absolute temperature, and $V^{ext}$ is the external potential.
In this work, we consider hydrogen adsorption in MOF structures where the external potential is mainly contributed by van der Waals interactions.
The Lennard-Jones 12-6 potential $V_{LJ}$ is therefore used to describe the pairwise additive potential for the interaction of a hydrogen molecule with each atom inside the MOF structure with the cutoff distance of 12.9 $\mathring{A}$.
\begin{equation}
{V_{LJ}} = 4\varepsilon \left[{\left(\frac{\sigma }{r }\right)^{12}} - {\left(\frac{\sigma }{r }\right)^6}\right]
\end{equation}
where $r$ is the pairwise distance between gas molecules and atoms in the MOFs, and $\sigma$ and $\varepsilon$ are the LJ size and energy parameters.
In this work, universal force field (UFF) is used for all atoms in MOFs.
For hydrogen molecule, ${\sigma}_{H_2}=0.296$ nm and ${\varepsilon}_{H_2 }/k_B=34.2$ K.
 For interactions between different kinds of atoms, the Lorentz-Berthelot mixing rules were used.

\section{The removal of MOF structures with extreme values} \label{appendix:sample}
\setcounter{figure}{0}
\begin{figure}[h]
	\centering
	\includegraphics[scale=0.5]{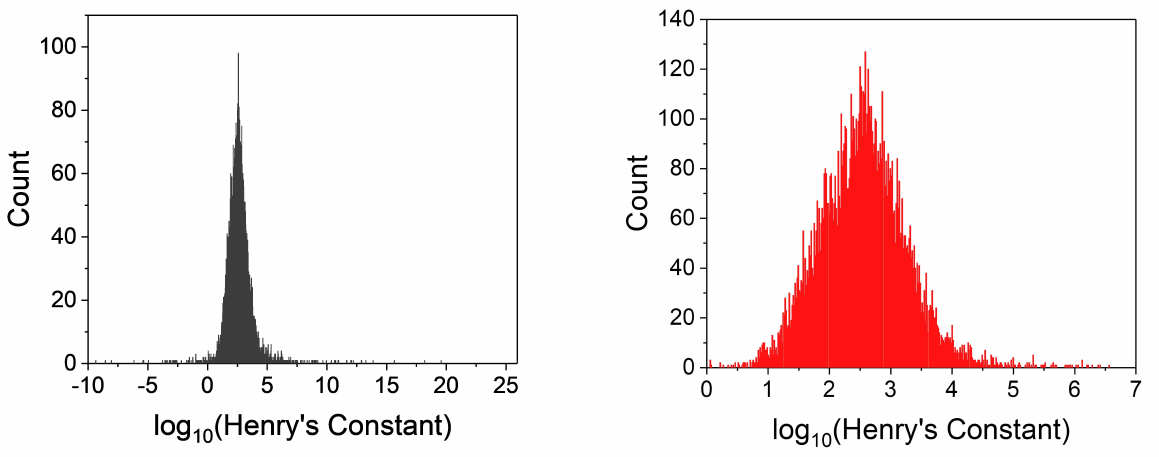}
	\caption{Distribution of Henry's constants for hydrogen gas in 12763 MOFs in original CoRE2019 MOF database (left) and 10136 MOFs after cleanup (right).}
	\label{henry1}
\end{figure}

The original CoRE2019 MOF database contains 12763 MOF structures.
According to our calculation of Henry's constant, the range of the Henry's constant covers more than 30 orders of magnitude (as shown in Figure \ref{henry1}).
After examining the MOF structure files, we found the broad range of Henry's constants was due to some unreasonable MOF structures.
In Figure \ref{henry1}, we can see that most MOFs distribute around unit Henry's constant after database cleanup.
There are a small portion of MOFs with exceptionally high Henry's constants in Figure \ref{henry1}.

A closer examination reveals that in these structures, atoms are unreasonably close to each other.
For example, in MOF (RUBLEH) with the highest Henry's constant, the distance of oxygen-oxygen there is only $0.366$ $\mathring{A}$, which is even shorter than the shortest covalent bond (hydrogen-hydrogen $0.74$ $\mathring{A}$). The overlapped atom position would lead to unreasonably high attraction and unreasonably high Henry's constant.

\begin{figure}[H]
	\centering
	\includegraphics[scale=.7]{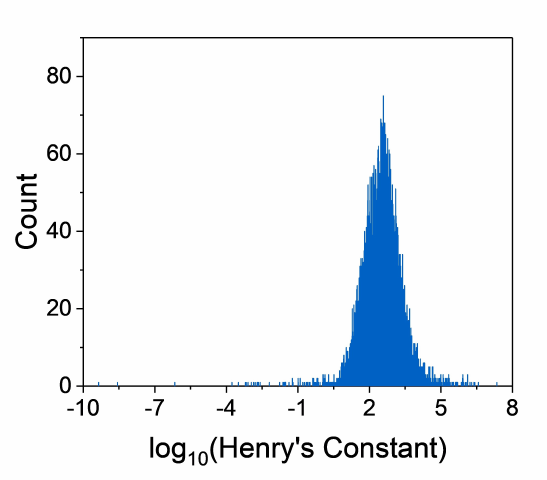}
	\caption{Distribution of Henry’s constant for MOFs without potentially overlapping atoms in CoRE2019 MOF database.}
	\label{henry2}
\end{figure}

After getting rid of MOF structure with potentially overlapped atoms (with smallest atom distance smaller than $0.8$ $\mathring{A}$), the distribution of Henry's constant is more reasonable as shown in Figure \ref{henry2}. Most MOFs with exceptionally high Henry's constant are filtered out.

\begin{figure}[H]
	\centering
	\includegraphics[scale=.7]{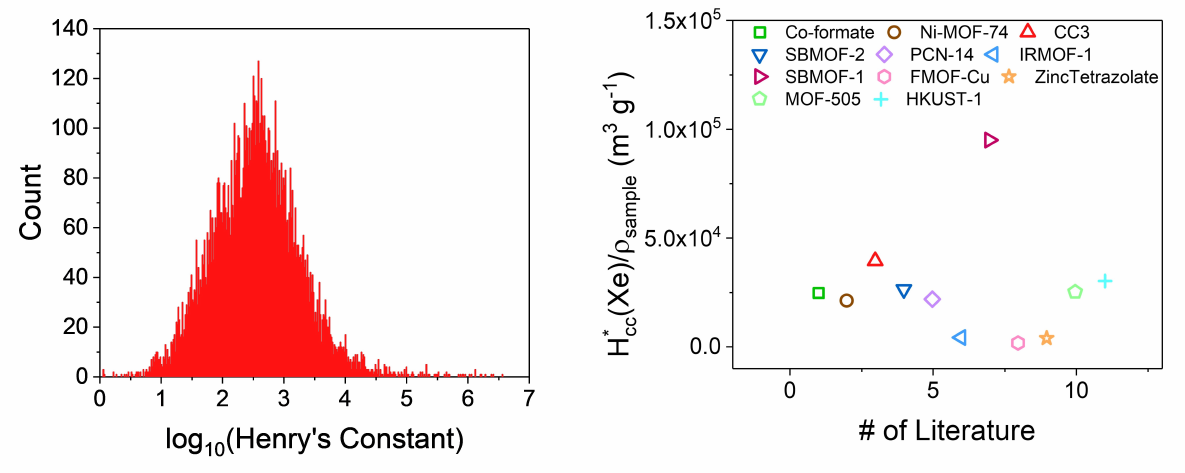}
	\caption{Distribution of Henry’s constants for hydrogen gas in MOFs without structures containing potentially overlapped atoms and extremely small surface area (left). Xe/Kr selectivity for Henry’s constant from the literature (right).}
	\label{henry3}
\end{figure}

It is also worth noticing that some Henry's constant is also extremely low (on the order of $10^{-10}$).
For the low Henry's constant for those MOFs is due to the relatively small pore size or tight confined geometry (XEKUO).
Compared to other MOFs with similar void fractions, XEHUO has a relatively thin structure and therefore most interactions inside the MOF are repulsive which results in an ultra low Henry's constant.
For example, compared with XEJJOM (void volume of 253 \AA$^3$), XEKUO (void volume of 185 \AA$^3$) has 0 surface area while XEJJOM has surface of 9.40 \AA$^2$ when using test particle with diameter of 3.68 \AA (size of nitrogen molecule).
After removing MOFs with surface areas less than 50 \AA$^2$ (meaning that there will be very little space and very unlikely for surface adsorption to happen), we filtered out all the MOFs with extremely low Henry's constant.
The cleaned data set contains 10136 MOF structures.
The Henry's constants still cover the several order of magnitudes.
However, we believe they may be reachable. The distribution of Henry's constants of the cleaned dataset is shown in Figure \ref{henry3}.

\end{document}